# Electrical control of a long-lived spin qubit in a Si/SiGe quantum dot

E. Kawakami[1†], P. Scarlino[1†], D. R. Ward[2], F. R. Braakman[1,3], D. E. Savage[2], M. G. Lagally[2], Mark Friesen[2], S. N. Coppersmith[2], M. A. Eriksson[2], and L. M. K. Vandersypen[1*]

[1]Kavli Institute of Nanoscience, TU Delft, Lorentzweg 1, 2628 CJ Delft, The Netherlands.

[2]University of Wisconsin-Madison, Madison, WI 53706, USA.

[3]Current address: Department of Physics, University of Basel, 4056 Basel, Switzerland

* Correspondence to: l.m.k.vandersypen@tudelft.nl

† Contributed equally

**Nanofabricated quantum bits permit large-scale integration but usually suffer from short coherence times due to interactions with their solid-state environment [1]. The outstanding challenge is to engineer the environment so that it minimally affects the qubit, but still allows qubit control and scalability. Here we demonstrate a long-lived single-electron spin qubit in a Si/SiGe quantum dot with all-electrical two-axis control. The spin is driven by resonant microwave electric fields in a transverse magnetic field gradient from a local micromagnet [2,3], and the spin state is read out in single-shot mode [4]. Electron spin resonance occurs at two closely spaced frequencies, which we attribute to two valley states. Thanks to the weak hyperfine coupling in silicon, Ramsey and Hahn echo decay timescales of $1\mu s$ and $40\mu s$, respectively, are observed. This is almost two orders of magnitude longer than the intrinsic timescales in III-V quantum dots [5,6], while gate operation times are comparable to those achieved in GaAs [3,7,8]. This places the single-qubit rotations in the fault-tolerant regime [9] and strongly raises the prospects of quantum information processing based on quantum dots.**

The proposal by Loss and DiVincenzo [10] to define quantum bits by the state of a single electron spin in a gate-defined semiconductor quantum dot has guided research for the past 15 years [7]. Most progress was made in well-controlled III-V quantum dots, where spin manipulation with two [6,11], three [12] and four [13] dots has been realized, but gate fidelities and spin coherence times are limited by the unavoidable interaction with the fluctuating nuclear spins in the host substrate [5,6]. While the randomness of the nuclear spin bath could be mitigated to some extent by feedback techniques [14], eliminating the nuclear spins by using group IV host materials offers the potential for extremely long electron spin coherence times that exceed one second in P impurities in bulk $^{28}Si$ [15,16].

Much effort has been made to develop stable spin qubits in quantum dots defined in carbon nanotubes [17,18], Ge/Si core/shell nanowires [19], Si MOSFETs [20,21] and Si/SiGe 2D electron gases [16,22,23]. However, coherent control in these group IV quantum dots is so far limited to a Si/SiGe singlet-triplet qubit with only single-axis control [23] and a carbon nanotube single-electron spin qubit, with a Hahn echo decay time of 65 ns [17].

Our device is based on an undoped Si/SiGe heterostructure with two layers of electrostatic gates (Fig. 1a). Compared to conventional, doped heterostructures, this technology strongly improves charge stability [23]. First, accumulation gates ($V_a \sim +150\,\mathrm{mV}$) are used to induce a two-dimensional electron gas (2DEG) in a 12 nm wide Si quantum well 37 nm below the surface. Second, a set of depletion gates, labelled 1-12 in Fig. 1a, is used to form a single or double quantum dot in the 2DEG, flanked by a quantum point contact and another dot intended as charge sensors. Two $1\,\mu\mathrm{m}$-wide, 200 nm-thick, and $1.5\,\mu\mathrm{m}$-long Co magnets are placed on top of the accumulation gates (Fig. 1a), providing a stray magnetic field with components $B_{//}$ and $B_{\perp}$, parallel and perpendicular to the external magnetic field, respectively. The sample is attached to the mixing chamber of a dilution refrigerator with base temperature ~25 mK and the electron temperature estimated from transport measurements is ~150 mK. We tune the right dot to the few-electron regime (Supplementary Fig. 1c) and adjust the tunnel rate between the dot and the reservoir to ~1 kHz, so that dot-reservoir tunnel events can be monitored in real time using the sensing dot (Fig. 1a). The left dot is not used in the experiment and the constrictions between gates 4 and 8 and between 3 and 10 are pinched off. Gates 3, 8, 9 and 11 are connected to high-frequency lines via bias-tees. Microwave excitation applied to one of these gates oscillates the electron wave function back and forth in the dot. Because of the gradient $dB_{\perp}/dx \sim 0.3$ mT/nm (Fig. 1b), the electron is then subject to an oscillating magnetic field. Electric dipole spin resonance (EDSR) occurs when the microwave frequency $f_{MW}$ matches the electron spin precession frequency in the magnetic field at the dot position [2,3].

All measurements shown here use 4-stage voltage pulses applied to gate 3 (Fig. 1c): (1) initialization to spin-down (4 ms, fidelity ~ 95% (Supplementary Section S7)), (2) spin manipulation through microwave excitation of gate 8 (1 ms), (3) single-shot spin read-out (4 ms, fidelity ~ 95% (Supplementary Section S7)), and (4) a compensation/empty stage (1 ms). By repeating this cycle, typically 150-1000 times, we obtain statistics of how often an electron leaves the dot during the detection stage, giving the spin-up probability $P_{\uparrow}$ at the end of the manipulation stage. The measured spin resonance frequency as a function of applied magnetic field is shown in Fig. 2a. We can extract the electron $g$-factor using the relation:
$$hf_0 = g\mu_B B_{\mathrm{local}}, \quad (1)$$
where $B_{\mathrm{local}} = \sqrt{(B_{\mathrm{ext}} + B_{//})^2 + B_{\perp}^2}$, $h$ is Planck's constant and $\mu_B$ is the Bohr magneton. From fits to equation. (1) (blue curve in Fig. 2a), we find $g = 1.998 \pm 0.002$, where we used $B_{//} = 120\,\mathrm{mT}$ and $B_{\perp} = 50\,\mathrm{mT}$, based on numerical simulation of the stray magnetic field from the micromagnet at the estimated dot location (Supplementary Section S2).

Surprisingly, when measuring the EDSR peak at a sufficiently low power to avoid power broadening, we resolve two lines, separated by 2 - 4 MHz in the range $B_{ext} = 0.55 - 1.2\,\text{T}$ (Fig. 2b). We return to the origin of this splitting later. Fitting each resonance peak with a Gaussian function yields $\delta f^{(2)}_{FWHM} = 0.63 \pm 0.06\,\text{MHz}$ for the higher-energy transition at frequency $f_0^{(2)}$ and $\delta f^{(1)}_{FWHM} = 0.59 \pm 0.05\,\text{MHz}$ for the lower-energy transition at frequency $f_0^{(1)}$. From this line width, we extract a dephasing time $T_2^* = \dfrac{\sqrt{2}\hbar}{g\mu_B\sigma_B} = \dfrac{2\sqrt{\ln 2}}{\pi\delta f_{FWHM}} = 840 \pm 70\,\text{ns}$ [7], 30-100 times longer than $T_2^*$ in III-V dots [6,5,7,8], and several times longer than the $T_2^*$ measured before in Si/SiGe dots [21,23]. This dephasing timescale can be attributed to the random nuclear field from the 5% $^{29}$Si atoms in the substrate with standard deviation $\sigma_B = 9.6\,\mu\text{T}$, consistent with theory [24]. Given the presence of a magnetic field gradient $dB_{\parallel}/dx \sim 0.2\,\text{mT/nm}$, the line width also gives an upper bound on the electron micromotion induced by low-frequency charge noise of ~ 50 pm (rms).

Coherent control of the electron spin is achieved by applying short high-power microwave bursts of duration $t_p$. Figure 3a shows the measured spin-up probability, $P_\uparrow$, as a function of $f_{MW}$ and burst time $t_p$, which exhibits the chevron pattern that is characteristic of high-quality oscillations. On resonance, the spin rotates at the bare Rabi frequency, $f_1$. When detuned away from resonance by $\Delta f = f_{MW} - f_0$, the spin rotates about a tilted axis, the oscillation frequency increases as $\sqrt{\Delta f^2 + f_1^2}$, and the visibility is reduced. The fast Fourier transform over the microwave burst time of the data in Fig. 3a is shown in Fig. 3c and exhibits the expected hyperbolic dependence as a function of $\Delta f$ for both transitions, $f_0^{(1)}$ and $f_0^{(2)}$. We fit both hyperbolae with one free parameter $f_1$ each (black rectangles and red circles), giving $f_1^{(1)} = 5.0 \pm 0.6\,\text{MHz}$ ($B_1 \sim 0.18\,\text{mT}$) and $f_1^{(2)} = 3.1 \pm 0.6\,\text{MHz}$ (errors arise from the finite number of points in the FFT) for the respective transitions. These single-spin Rabi frequencies are comparable to those observed in GaAs [3,8]. The relative amplitude of the oscillations at $f_0^{(1)}$ and $f_0^{(2)}$ is about 30/70; note that despite its lower weight, the peak at $f_0^{(1)}$ is tallest in Fig. 2b, since its Rabi frequency is a factor of 1.5 ± 0.2 higher than that of the other peak (Supplementary Section S5). The extracted Rabi frequencies of both transitions are proportional to the microwave amplitude, as expected (Supplementary Fig. 6).

The observed decay of the Rabi oscillations cannot be explained only by the spread in the Larmor frequency, $\sigma_B$. Numerical simulations of the Rabi oscillations give good agreement with the measurements of Fig. 3a when including a variation in the Rabi frequency, $\sigma_{Rabi} \sim 0.25\,\text{MHz}$ (Supplementary Section S4). The fluctuations in the transverse nuclear field [25] are too small to explain this spread. Instead, instrumentation noise could be responsible. Taking into account $f_1$, $\sigma_B$, and $\sigma_{Rabi}$, we estimate a gate fidelity for a $\pi$ rotation of 0.99 (0.97) for an electron spin resonant at $f_1^{(1)}$ ($f_1^{(2)}$). For an electron in a 30/70 statistical mixture of the two resonance conditions, the fidelity is ~0.80 (Supplementary Section S7).

Two-axis control of the spin is demonstrated by varying the relative phase $\phi$ of two $\pi/2$ microwave bursts resonant with $f_1^{(2)}$ separated by a fixed waiting time $\tau = 40$ ns $\ll T_2^*$ (Fig. 3f, black trace). As expected, the signal oscillates sinusoidally in $\phi$ with period $2\pi$. For $\tau = 2$ μs $> T_2^*$, the contrast has vanished, indicating that all phase information is lost during the waiting time (Fig. 3f, red trace). Similar measurements with the pulses applied off-resonance by an amount $\Delta f$ with $\phi = 0$, are expected to show an oscillation with frequency $\Delta f$ and an envelope that decays on the timescale $T_2^*$. Because of the presence of two resonance lines just 2.1 MHz apart, the measurement of $P_\uparrow$ versus $f_{MW}$ and $\tau$ (Fig. 3b) shows a superposition of two such patterns. This becomes clear from taking the Fourier transform over the waiting time $\tau$ (Fig. 3d) which shows 2 linear patterns superimposed, with vertices at $f_0^{(1)}$ and $f_0^{(2)}$. The stability of the measurement can be appreciated from Fig. 3e, which shows $P_\uparrow$ versus $f_{MW}$ and the relative phase between the two bursts at $\tau = 400$ ns.

Spin coherence can be extended by spin echo techniques, provided the source of dephasing fluctuates slowly on the timescale of the electron spin dynamics. We perform a Hahn echo experiment, consisting of $\pi/2$, $\pi$ and $\pi/2$ pulses separated by waiting times $\tau/2$ [7,16], and record $P_\uparrow$ as a function of the total free evolution time $\tau$ (Fig. 4a). A fit to a single exponential yields a time constant $T_2 = 37 \pm 3$ μs, almost 50 times longer than $T_2^*$, and 50-100 times longer than the Hahn echo decay time in GaAs dots [6,26]. The decay is well-described by a single exponential, with no signatures of a flat top. This indicates that the fluctuations that dominate the echo decay are fast compared to the few μs timescale of the first few data points [27]. That is consistent with the fact that a four-pulse decoupling pulse sequence does not further extend the decay time (Fig. 4b). This implies that the slowly fluctuating nuclear field does not yet limit $T_2$ [28]; indeed, a 200 μs Hahn echo decay was observed for an electron spin bound to P impurity in silicon [29]. Jumping between the two resonance conditions on a few μs timescale is not likely to be responsible for the decay either, since the line width implies that the resonance condition is stable on a timescale of at least 1 μs. Presumably instrumentation or charge noise is dominant. Finally, when we shift the position of the third pulse, the time intervals before and after the echo pulse are no longer equal and coherence is lost, as expected (Fig. 4c). A fit of this decay with a Gaussian function, gives $T_2^* = 920 \pm 70$ ns measured in the time domain, consistent with $T_2^*$ extracted from the line width.

We now return to the origin of the two resonance lines that are visible in all the measurements. From the individual measurements, we deduce that the higher (lower) frequency resonance contributes to the signal 70% (30%) of the time, indicating that the system does not simply exhibit two resonances but instead switches between two conditions. Since the switches do not dominate the Hahn echo decay, they must be slower than ~ 50 μs. The splitting between the two lines varies linearly with $B_{ext}$, corresponding to a difference in *g*-factors of about 0.015%, and an offset in $B_{local}$ between the two resonances of $65 \pm 138$ mT (Fig. 2a, green triangles). Finally, as mentioned before, the higher-frequency resonance exhibits ~ 1.5 times slower Rabi oscillations than the lower-frequency resonance.

We propose that the two lines correspond to EDSR with the electron in one or the other of the two lowest valley states, with a 30/70 occupation ratio. This ratio is set either by the injection probabilities into the respective valley states, or by thermal equilibration, depending on whether the valley lifetime is shorter than the few ms delay between injection and manipulation. We note that either way, initialization to a single valley can be achieved when the valley splitting is several times larger than the electron temperature. A valley-dependent spin splitting can arise from several sources. Intrinsic spin-orbit coupling is weak in silicon, but the field gradient from the micromagnet admixes spin and orbitals, leading to a renormalization of the $g$-factor by an amount that depends on the orbital level spacing [2]. Due to valley-orbit coupling, the orbital level spacing in turn depends on the valley. We estimate that this can result in observed valley-dependent $g$-factor shifts of ~ 0.015% (Supplementary Section S10). The difference in Rabi frequencies can be understood from a valley-dependent orbital level spacing as well [3]. Another mechanism that can account for the observed $g$-factor shifts is valley-dependent penetration of the Bloch wave function into the SiGe barrier region (Supplementary Section S10). Other explanations we considered include switching between two separate dot locations, a double dot, and transitions in a two-electron manifold, but these are not consistent with the above observations; see also the supplementary information.

The demonstration of all-electrical single-spin control with coherence times orders of magnitude longer than intrinsic coherence timescales in III-V hosts greatly enhances the promise of quantum dot based quantum computation. Single qubit gate fidelities estimated for current parameters with valley splitting control [20], reach the fault-tolerance threshold of the recently developed surface codes [9]. The use of a micromagnet facilitates selective addressing of neighbouring spins and provides a coupling mechanism of quantum dot spins to stripline resonators that can form the basis for two-qubit gates and a scalable architecture [30].

**FIGURE CAPTIONS:**

**Figure 1: Device schematic and measurement cycle.**
**a**, False color device image showing a fabricated pattern of split gates, labeled 1-12. For this experiment we create a single quantum dot (QD), estimated location at the red circle) and a sensing dot (SD). The current $I_{SD}$ is measured as a function of time for a fixed voltage bias $V_{SD} = -600\,\mu\text{V}$. The voltage pulses are applied to gate 3 and the microwaves are applied to gate 8. The blue semi-transparent rectangles show the position of two Co 200 nm thick micromagnets. The yellow shaded pieces show the location of two accumulation gates, one for the reservoirs and another for the double quantum dot region.
**b**, Numerically computed magnetic field component perpendicular to the external field, induced by the micromagnet in the plane of the Si quantum well, for fully magnetized micromagnets. The straight solid lines indicate the edges of the micromagnet as simulated. The region shown is outlined with dotted lines in panel (A).
**c**, Microwave (MW) and gate voltage pulse scheme (see main text) along with an example trace of $I_{SD}$ recorded during the pulse cycle and cartoons illustrating the dot alignment and tunnel events. During stages (1) and (3), the Fermi level in the reservoir is set in between the spin-down and spin-up energy levels so that only a spin-down electron can tunnel into the dot and only a spin up electron can tunnel out [4]. During stage (2), the dot is pulsed deep into Coulomb blockade, in order to minimize photon-assisted tunneling. The MW burst of duration $t_p$ ends about 100 ns—500 ns before the detection stage. When a step is observed during stage (3), see the dotted line, we count the electron as spin-up. Stage (4) serves to keep the DC component of the pulse zero and to symmetrize pulse distortions from the bias tee. In the process, the QD is emptied. The spike during the manipulation stage is due to the influence of the microwave burst (here 700 µs) on the detector.

**Figure 2: Qubit spectroscopy**
**a**, Measured microwave frequency that matches the electric dipole spin resonance (EDSR) condition $f_0^{(1)}$ (dark blue and light blue circles) and the difference between the two resonance frequencies $f_0^{(2)} - f_0^{(1)}$ (green triangles) as a function of externally applied magnetic field. The 6 points where $f_{MW} > 20\,\text{GHz}$ are measured by two-photon transitions [25]. The microwave burst time $t_p = 700\,\mu s \gg T_2^*$, effectively corresponding to continuous wave (CW) excitation (here we used low power excitation, P = -33 dBm to -10 dBm at the source, decreasing with lower microwave frequency). The upper of the two resonances in panel (b) is shown. The blue solid curve is a fit to the dark blue circles using equation (1). The light blue circles are excluded from the fit; presumably the micromagnet begins to demagnetize here. The green line is a linear fit to the green triangles.
**b**, Measured spin-up probability $P_\uparrow$ as a function of applied microwave frequency $f_{MW}$ for $B_{ext} = 560.783\,\text{mT}$ (P = -33 dBm), averaged over 200 minutes, i.e. 1 200 000 single-shot measurements.

**Figure 3: Universal qubit control**
**a**, Measured spin-up probability, $P_\uparrow$, as a function of $f_{MW}$ and burst time $t_p$ ($B_{ext} = 560.783\,\text{mT}$, $P = 16.4\text{dBm}$).
**b**, Measured spin-up probability, $P_\uparrow$, as a function of $f_{MW}$ and waiting time $\tau$ ($B_{ext} = 560.783\,\text{mT}$, $P = 16.4\text{dBm}$) between two $\frac{\pi}{2}$ (75 ns) pulses with equal phase, showing Ramsey interference (Color map as in (a)).
**c**, Fourier transform over the microwave burst time $t_p$ of Fig. 3A showing a hyperbolic dependence (black rectangles and red circles) as a function of $f_{MW}$ for each transition, $f_0^{(1)}$ and $f_0^{(2)}$. Inset: microwave pulse scheme used in (a).
**d**, Fourier transform over the waiting time $\tau$ of Fig. 3b showing two linear patterns superimposed, with vertices at $f_0^{(1)}$ and $f_0^{(2)}$. Inset: microwave pulse scheme used in (b,d,e,f). (Color map as in (c))
**e**, Measured spin-up probability, $P_\uparrow$, as a function of $f_{MW}$ and the relative phase $\phi$ between two microwave pulses for $\tau = 400\,\text{ns}$ ($B_{ext} = 763.287\text{mT}$, $P = 18.8\text{dBm}$). (Color map as in (a))

**f**, Ramsey signal as a function of the relative phase $\phi$ between the two microwave pulses for $\tau = 40\,\text{ns}$ (black curves) and $\tau = 2\,\mu\text{s}$ (red curves) ($B_{\text{ext}} = 763.287\,\text{mT}$, $P = 18.8\,\text{dBm}$, $f_{MW} = f_0^{(2)} = 18.41608\,\text{GHz}$).

### Figure 4: Qubit coherence

(Here, $B_{\text{ext}} = 747.710\,\text{mT}$, $P = 18.4\,\text{dBm}$, $f_{MW} = f_0^{(2)} = 17.695\,\text{GHz}$; $f_1^{(2)} = 2.7\,\text{MHz}$.)

**a**, Measured spin-up probability as a function of the total free evolution time $\tau$ in a Hahn echo experiment (pulse scheme in inset).

We did not see a significant difference in the decay when changing the relative phase between the first pulse (77 ns) and the $\pi$ pulse (150 ns) from $\phi = 0$ to $\phi = 90$. The decay curve is fit well to a single exponential decay.

**b**, Measured spin-up probability as a function of the total free evolution time $\tau$ when using four decoupling pulses.

**c**, Measured spin-up probability as a function of the position of the third pulse in the Hahn echo experiment.

The free evolution time between the first and second pulse is fixed at $5\,\mu\text{s}$ and that between the second and third pulse is varied from 3 to $7\,\mu\text{s}$.

**Acknowledgments:** We acknowledge useful discussions with L. Schreiber, J. Prance, G. de Lange, W. Coish, F. Beaudoin, and our spin qubit teams, comments by L. DiCarlo and R. Hanson, and experimental assistance by P. Barthelemy, M.Tiggelman, and R. Schouten. This work was supported in part by ARO (W911NF-12-0607), FOM and the ERC; development and maintenance of the growth facilities used for fabricating samples is supported by DOE (DE-FG02-03ER46028). E.K. was supported by a fellowship from the Nakajima Foundation. This research utilized NSF-supported shared facilities at the University of Wisconsin-Madison.

**Author contributions:**

E.K. and P.S performed the experiment with help from F.R.B., and analyzed the data, D.R.W. fabricated the sample, D.E.S and M.G.L. grew the heterostructure, E.K., P.S., M.F, S.N.C., M.A.E. and L.M.K.V. carried out the interpretation of the data, and M.F and S.N.C. the theoretical analysis. E.K., P.S., and L.M.K.V. wrote the manuscript and all authors commented on the manuscript. M.A.E. and L.M.K.V. initiated the project, and supervised the work with S.N.C.

**Additional information:**
Supplementary information is available in the online version of the paper. Reprints and permissions information is available online at www.nature.com/reprints.

**Competing financial interests:**
The authors declare no competing financial interests.

# Figure 1

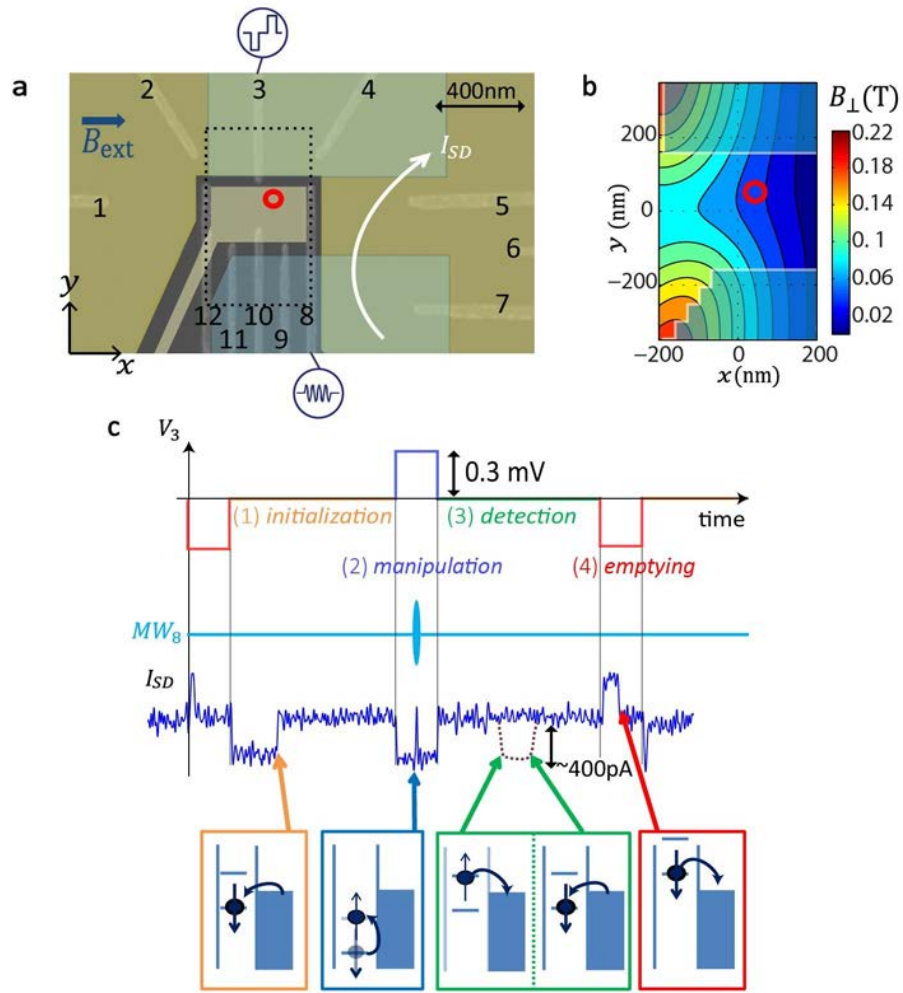

Figure 2

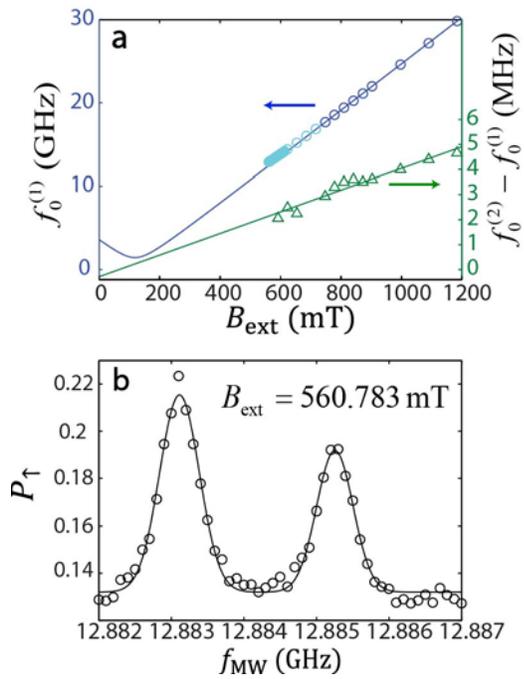

Figure 3

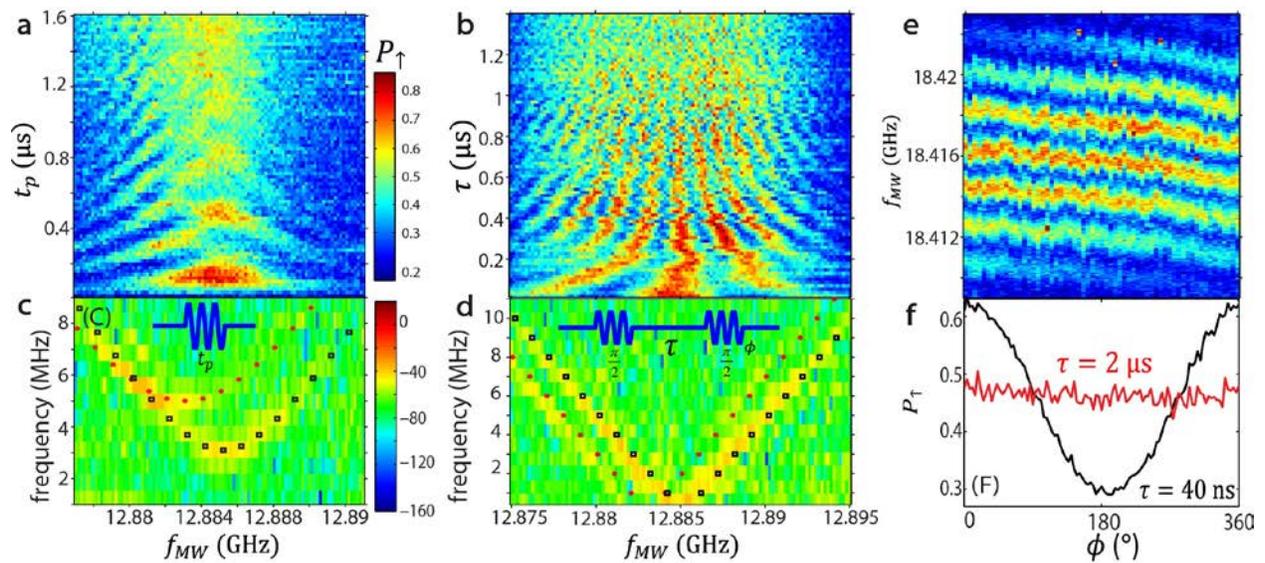

Figure 4

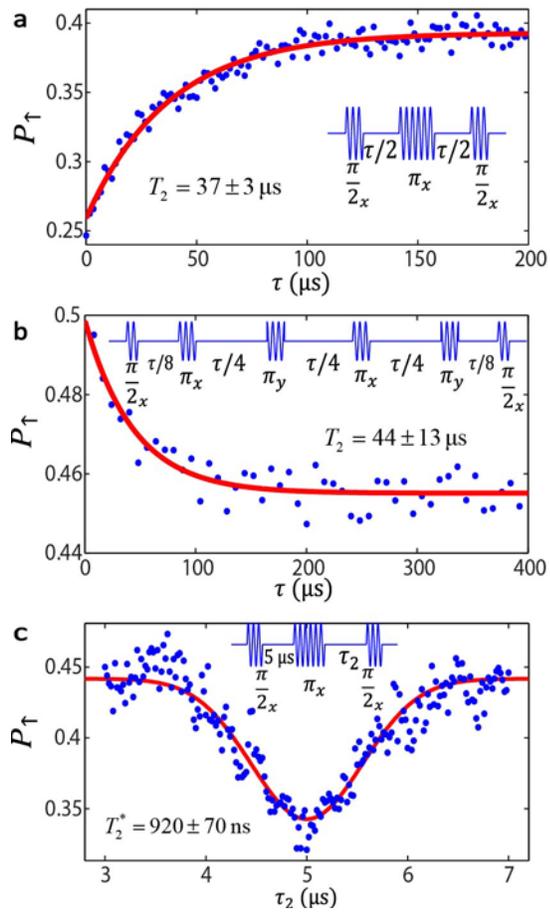

Supplementary Materials for

# Electrical control of a long-lived spin qubit in a Si/SiGe quantum dot

E. Kawakami[†], P. Scarlino[†], D. R. Ward, F. R. Braakman, D. E. Savage, M. G. Lagally, Mark Friesen, S. N. Coppersmith, M. A. Eriksson, and L. M. K. Vandersypen[*]

* Correspondence to: l.m.k.vandersypen@tudelft.nl

† Contributed equally

**Supplementary Materials:**

Supplementary text (Sections 1-10)

Supplementary figures. S1 to S8

References

**Summary:**

      S1. Sample fabrication, characterization and measurement techniques
      S2. Magnetic field gradient induced by micromagnets
      S3. Spin relaxation time $T_1$
      S4. Rabi oscillation
      S5. Population and time dependence of the two resonances
      S6. π pulse fidelity
      S7. Initialization fidelity and readout fidelity
      S8. Power dependence of the Rabi frequency
      S9. Measurements of Ramsey fringes
      S10. Difference in *g*-factors and Rabi frequencies between the two resonances

# S1. Sample fabrication, characterization and measurement techniques

**Sample fabrication**

An SEM image of the sample used in this experiment is shown in Fig. S1A. While Fig. 1a shows only the SEM image of the depletion gates, Fig. S1A shows also the two accumulation gates and the two micro magnets. The epitaxial structure, shown in Fig. S1B, is grown by chemical vapor deposition. An 800 nm $Si_{0.7}Ge_{0.3}$ buffer is deposited on a substrate, followed by a 12 nm thick strained Si well. A 32 nm $Si_{0.7}Ge_{0.3}$ layer is then deposited, followed by a 1 nm thick Si cap layer. The sample is undoped; charge carriers are induced in the Si quantum well by application of positive voltages to the accumulation gates, forming a 2DE [1,2]. To minimize unwanted accumulation and charge leakage, most of the substrate is etched to below the Si quantum well using reactive ion etching, leaving active material for the dot structures only in small 100 $\mu m^2$ ×100 $\mu m^2$ mesas. All exposed surfaces are then uniformly coated with 10 nm of $Al_2O_3$ via atomic layer deposition (ALD). Ohmic contacts to the 2DEG are created by 20 kV phosphorus implantation activated with a 15 s, 700°C anneal. Two layers of gates, separated by an isolating layer of 80 nm of $Al_2O_3$ deposited by ALD, are defined by a combination of photo- and electron-beam lithography and deposited by electron-beam evaporation of 1.7 nm Ti/40 nm Au. Two Co micro magnets are defined on top of the upper layer of gates by electron-beam lithography and deposited by electron-beam evaporation of 5 nm Ti/200 nm Co/20 nm Au. The top Au layer minimizes oxidation of the Co material.

The sample is glued on a printed circuit board (PCB) with 4 high-frequency lines connected to gates 3, 8, 9 and 11. Those lines are fitted with homemade resistive bias tees on the PCB (R = 10 MΩ, C = 47 nF; 1/RC ~2 Hz) to allow fast pulsing of the gate voltages while also maintaining a DC bias on the gates. The presence of the bias tee is the reason why we use four stage pulses while we could have used two stage pulses [33]. The extra two stages make the voltage level during the initialization and detection stages much less variable. The high-frequency lines contain a 20 dB attenuator at 1 K and a 10 dB attenuator at the mixing chamber.

**Quantum dot characterization**

The right dot is tuned to the few-electron regime by adjusting the voltages on the gates 3, 4, 5, 8, 9 and 10. Fig. S1C shows the differential transconductance $\frac{dI_{SD}}{dV_{gate3}}$ as a function of the voltages on gates 3 and 5. No other charge transitions are observed when pushing the voltage of gate 3 down to -375 mV with the other gate voltages kept at the same values as used in Fig. S1C, which permits us to assign tentative absolute electron numbers as shown in Fig. S1C. The experiment is done at the 0-1 charge transition. This QD presents an addition energy of 9 meV and an orbital level spacing of 450 μeV, estimated by pulse spectroscopy measurements. From the addition energies we extract a dot radius of 21 nm (in the approximation of a circular QD); from the orbital level spacing we deduce 28 nm assuming a harmonic confining potential and again a circular dot. Pulse spectroscopy measurements (not reported here) also show the linear dependence of the Zeeman splitting of the lowest orbital state as a function of external magnetic field, allowing us to calibrate the conversion factor between pulse amplitude and energy.

**Charge detection**

Thanks to the capacitive coupling between the dot and the sensing QD, the current level of the sensing QD is decreased (increased) by ~400 pA when an electron jumps from the dot to the reservoir (from the reservoir to the dot). We use a room temperature IV converter to record the sensing dot current, $I_{SD}$, using a low-pass filter with ~20 kHz cut-off to obtain a sufficient signal-to-noise ratio.

**Heating effects from the microwave bursts**

The application of high power microwave bursts affects the response of the sensing dot, presumably due to heating, and this effect increases with burst time. In order to keep the response constant and get better uniformity in the visibility of the spin oscillations as we vary the burst time during the manipulation stage, we include a second microwave burst at the end of the readout stage such that the total microwave burst duration over a full cycle is kept constant at 2 µs.

**Finding the spin resonance condition**

Before performing the experiment, the electron spin *g*-factor is not precisely known. The presence of the micromagnet creates further uncertainty in the spin resonance condition. The continuous wave low power EDSR response exhibits very narrow lines, making it easy to miss the resonance when scanning the magnetic field or frequency for the first time. At higher power, the line is power broadened, so larger steps in field or frequency can be taken, accelerating the scan. We used an even more efficient technique, adiabatic rapid passage. This technique was successfully used in quantum dots before[4] and allows one to step the frequency in increments corresponding to the frequency chirp range used for the adiabatic inversion (40-60 MHz in our experiments).

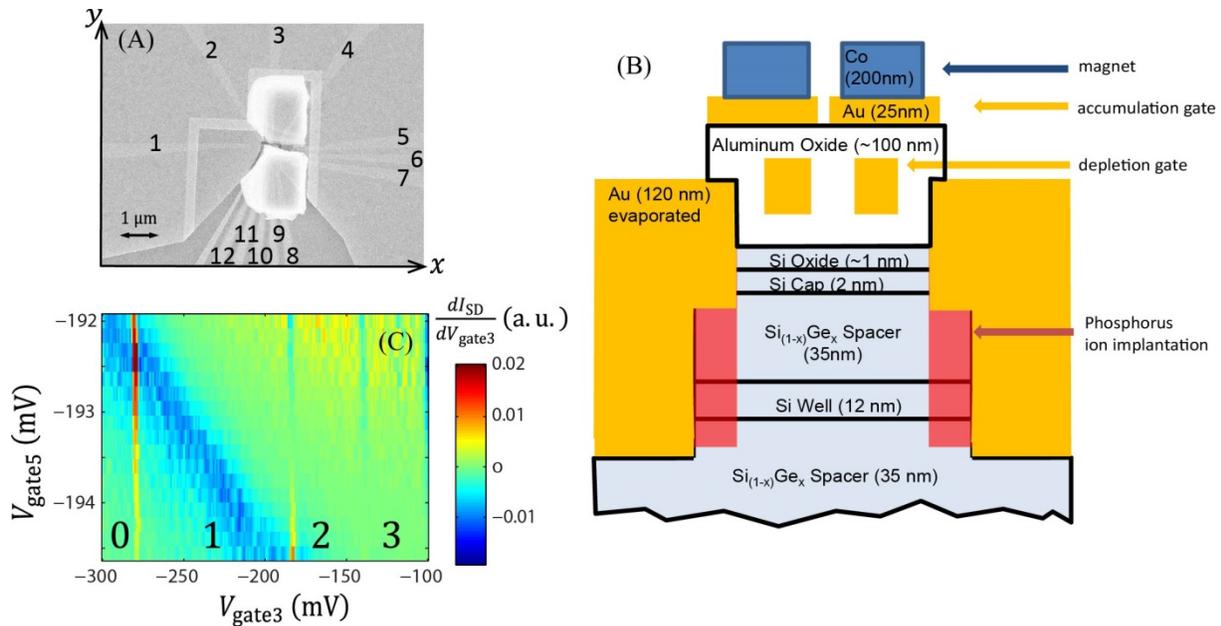

Fig. S1 Device schematic and charge stability diagram
(**A**) Scanning electron micrograph of the sample. The white regions around the area of the micro magnets are thin pieces of metal that were bent upwards during lift-off. (**B**) Schematic cross-section of the device. (**C**) Charge stability diagram of the single dot system, measured via the sensing quantum dot differential transconductance as a function of $V_{gate3}$ and $V_{gate5}$. The sharp nearly vertical lines correspond to changes in the dot occupation. The broad diagonal blue line corresponds to a Coulomb peak in the sensing dot. The tentative absolute electron numbers 0-3 are shown.

## S2. **Magnetic field gradient induced by micromagnets**

Figure S2 shows the result of a numerical calculation of the magnetic field created by the two micro magnets along the x, y and z directions, where z is perpendicular to the quantum well and x and y are marked in Fig. S1[5,6]. The external magnetic field is applied along x. From this simulation, we obtain the magnitudes of the magnetic field and the magnetic field gradient at the position of the dot of $B_\parallel = B_x = 120 \, \text{mT}$, $B_\perp = \sqrt{B_y^2 + B_z^2} = 50 \, \text{mT}$, $dB_\perp / dx \sim 0.3 \, \text{mT/nm}$, $dB_\perp / dy \sim 0.04 \, \text{mT/nm}$, $dB_\parallel / dx \sim 0.2 \, \text{mT/nm}$ and $dB_\parallel / dy \sim 0.05 \, \text{mT/nm}$.

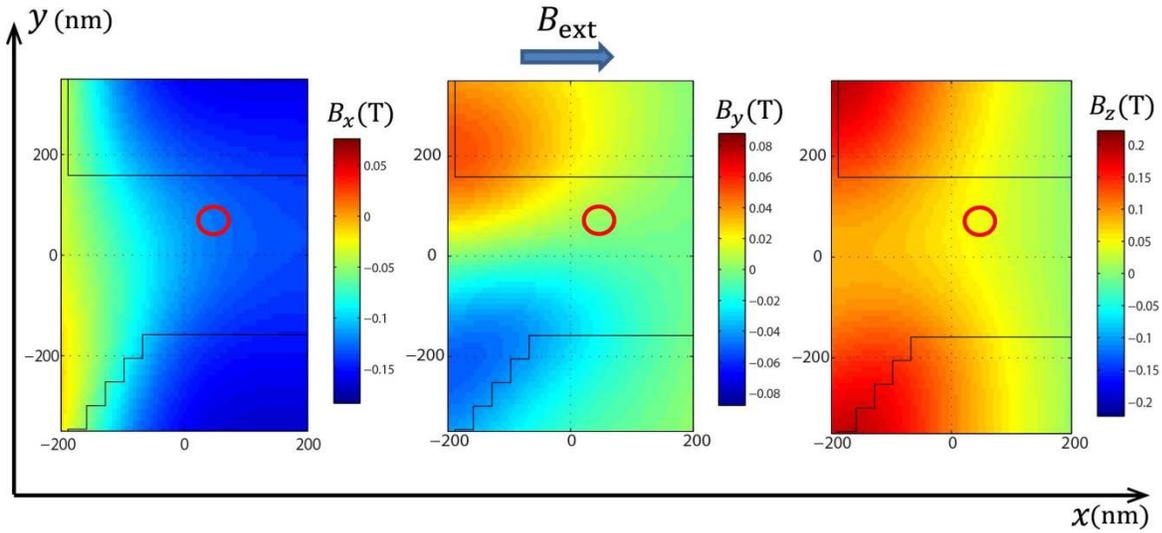

Fig. S2 Numerically computed x, y and z components of the magnetic field induced by the micromagnets in the plane of the Si quantum well, for fully magnetized micromagnets. The black solid lines indicate the edges of the micromagnet as simulated. The region shown is the same as that in Fig. 1b, and is outlined with dotted lines in Fig. 1a. The red circle shows the estimated position of the dot.

# S3. Spin relaxation time $T_1$

We did not observe any change in the measured spin-up probability when changing the timing of the microwave burst during the manipulation stage by up to 2 ms. Thus we conclude that the spin relaxation time $T_1$ is much longer than the ms timescale of our pulse cycle and that the measurements shown here are not affected by $T_1$ decay, consistent with the long $T_1$ times seen in earlier measurements on Si or Si/SiGe dots and donor 37,38,39.
We note that due to the very long spin relaxation time, we cannot initialize by equilibration, as was commonly done in previous work 40,41, since this would take 100 ms or more. Therefore, instead of pulsing both the spin up and spin down levels below the reservoir Fermi energy, thereby pulling an electron of unknown spin state inside the QD, we pulse so that only the lowest energy spin level (spin down) is below the Fermi level of the reservoir during the initialization stage (stage (1) in Fig. 1c of the main text).

## S4. Rabi oscillation

The probability that a single spin with Larmor frequency $f$ flips when it is subject to microwave excitation at frequency $f_{MW}$ with an amplitude that gives a Rabi frequency $f_1$ for a duration $t_p$ 33

$$p_{\uparrow\downarrow}(f_1, f, f_{MW}, t_p) = \sin^2\theta \sin^2\left(\pi t_p \sqrt{(f_{MW} - f)^2 + f_1^2}\right) \quad \text{Eq. (S1)}$$

$$\text{with } \sin\theta = \frac{f_1}{\sqrt{(f_{MW}-f)^2 + f_1^2}}.$$

There are two resonance frequencies, $f_0^{(1)}$ and $f_0^{(2)}$, as discussed in the main text. Here we assume that the populations in resonances (1) and (2) are $\varepsilon^{(1)}$, and $\varepsilon^{(2)}$, respectively. If in addition we assume that both the Larmor frequency and the Rabi frequency follow a (Gaussian) distribution, the spin flip probability is given by

$$P_{\uparrow\downarrow}(f_{MW}, t_p) = \sum_{n=1,2} \varepsilon^{(n)} P^{(n)}{}_{\uparrow\downarrow}(f_{MW}, t_p), \quad \text{Eq. (S2)}$$

with $P^{(n)}{}_{\uparrow\downarrow}(f_{MW}, t_p) = \int df_1 \int df\, G^{(n)}(f) g^{(n)}(f_1) p^{(n)}{}_{\uparrow\downarrow}(f_1, f, f_{MW}, t_p)$,

$$G^{(n)}(f) = \frac{1}{\sigma_f \sqrt{2\pi}} \exp\left(-\frac{\left(f - f_0^{(n)}\right)}{2\sigma_f^2}\right),$$

$$g^{(n)}(f_1) = \frac{1}{\sigma_{f_1}\sqrt{2\pi}} \exp\left(-\frac{(f_1 - f_1^{(n)})}{2\sigma_{f_1}}\right).$$

The standard deviation of the Larmor frequency $\sigma_f = 0.268 MHz$ is extracted directly from the line width (Fig. 2b).

In order to estimate the standard deviation of the Rabi frequency, $\sigma_{f_1}$, and the ratio of the two populations $\varepsilon^{(1)}:\varepsilon^{(2)}$ that applies in the experiment, we compare the measurement results of Fig. 3a with results from numerical simulations for $P_{\uparrow\downarrow}(f_{MW}, t_p)$ shown in Fig. S3A for a range of values for both the ratio $\varepsilon^{(1)}:\varepsilon^{(2)}$ and $\sigma_{f_1}$. Based on this rough comparison, we consider the agreement the best for $\varepsilon^{(1)}:\varepsilon^{(2)} \sim 0.3 \pm 0.1 : 0.7 \pm 0.1$ and $\sigma_{f_1} \sim 0.25 \pm 0.05$ MHz.

As a further consistency check, we plot the same simulation results again in Fig. S3B, but now rescaled to account for the read-out and initialization fidelities estimated in supplementary section 7 below. The same values $\varepsilon^{(1)}:\varepsilon^{(2)} \sim 0.3 \pm 0.1 : 0.7 \pm 0.1$ and $\sigma_{f_1} \sim 0.25 \pm 0.05$ MHz give good agreement with the data of Fig. 3a.

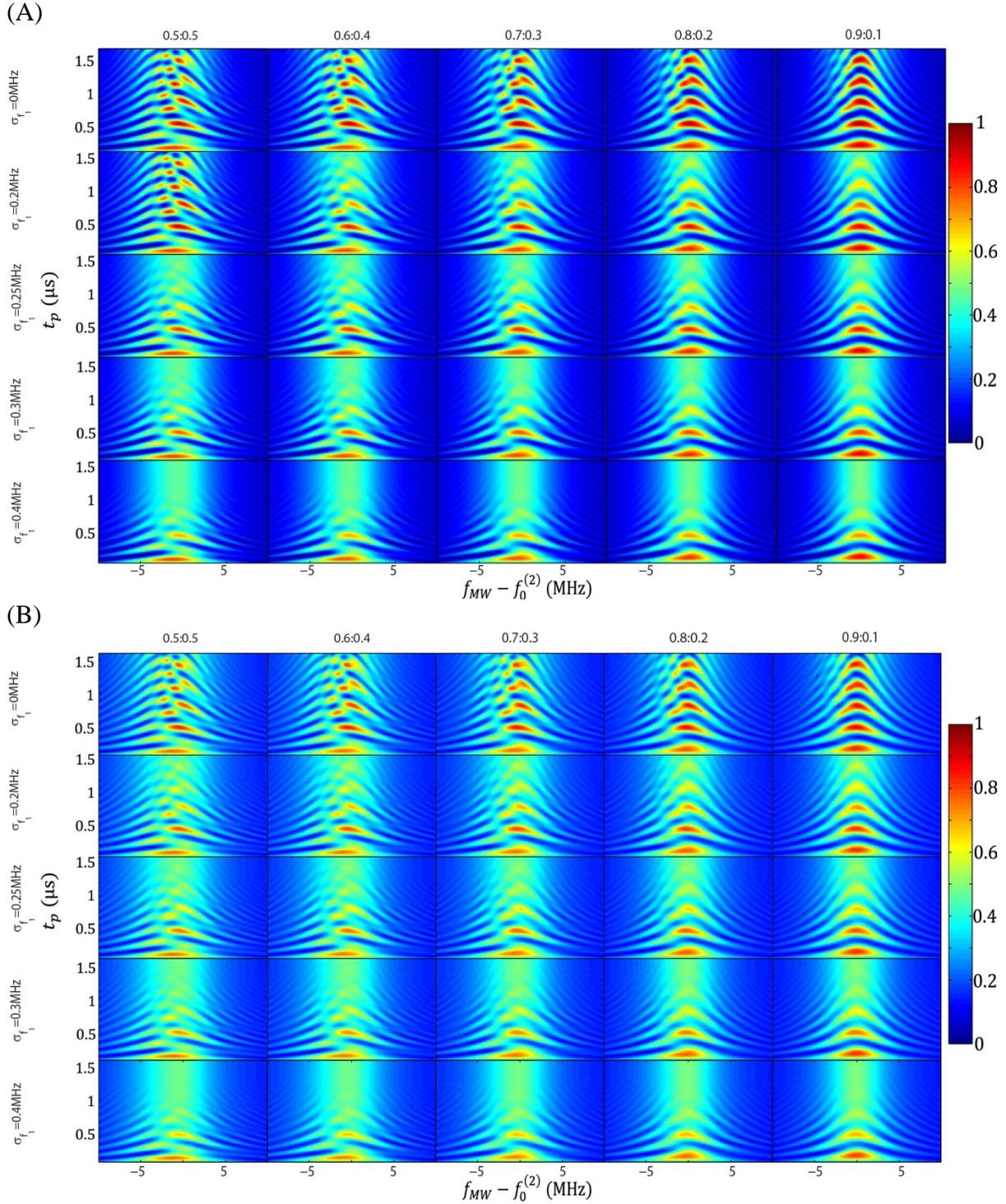

Fig. S3 Simulation for Rabi oscillations
(**A**) Numerically simulated spin flip probability $P_{\uparrow\downarrow}(f_{MW}, t_p)$ for population ratios $\varepsilon^{(1)}:\varepsilon^{(2)} = 0.5:0.5, 0.4:0.6, 0.3:0.7, 0.2:0.8, 0.1:0.9$ and Larmor frequency spread $\sigma_{f_1} = 0, 0.2$ MHz, $0.25$ MHz, $0.3$ MHz, $0.4$ MHz, as a function of driving duration $t_p$ and frequency detuning $f_{MW} - f_0^{(2)}$. From comparison with the data of Fig. 3a, we conclude that $\varepsilon^{(1)}:\varepsilon^{(2)} \sim 0.3 \pm 0.1 : 0.7 \pm 0.1$ and $\sigma_{f_1} \sim 0.25 \pm 0.05$ MHz are reasonable. (**B**) The same simulation results as in panel A, but taking into account the initialization and read-out fidelities estimated in section 7 below. Again $\varepsilon^{(1)}:\varepsilon^{(2)} \sim 0.3 \pm 0.1 : 0.7 \pm 0.1$ and $\sigma_{f_1} \sim 0.25 \pm 0.05$ MHz match well to the data.

Figure S4A shows the measured spin-up probability, $P_\uparrow$, as a function of $f_{MW}$ and burst time $t_p$, for $B_{ext} = 763.287$ mT. At this magnetic field, the two resonances are separated by $f_0^{(2)} - f_0^{(1)} = 2.838$ MHz. Thus the individual chevron patterns produced by two resonances are more easily distinguished than in Fig. 3a. The numerical simulations for $P_{\uparrow\downarrow}(f_{MW}, t_p)$ for $f_0^{(2)} - f_0^{(1)} = 2.838$ MHz taking into account the read-out and initialization fidelities estimated in supplementary section 7 are shown in Fig. S4B.

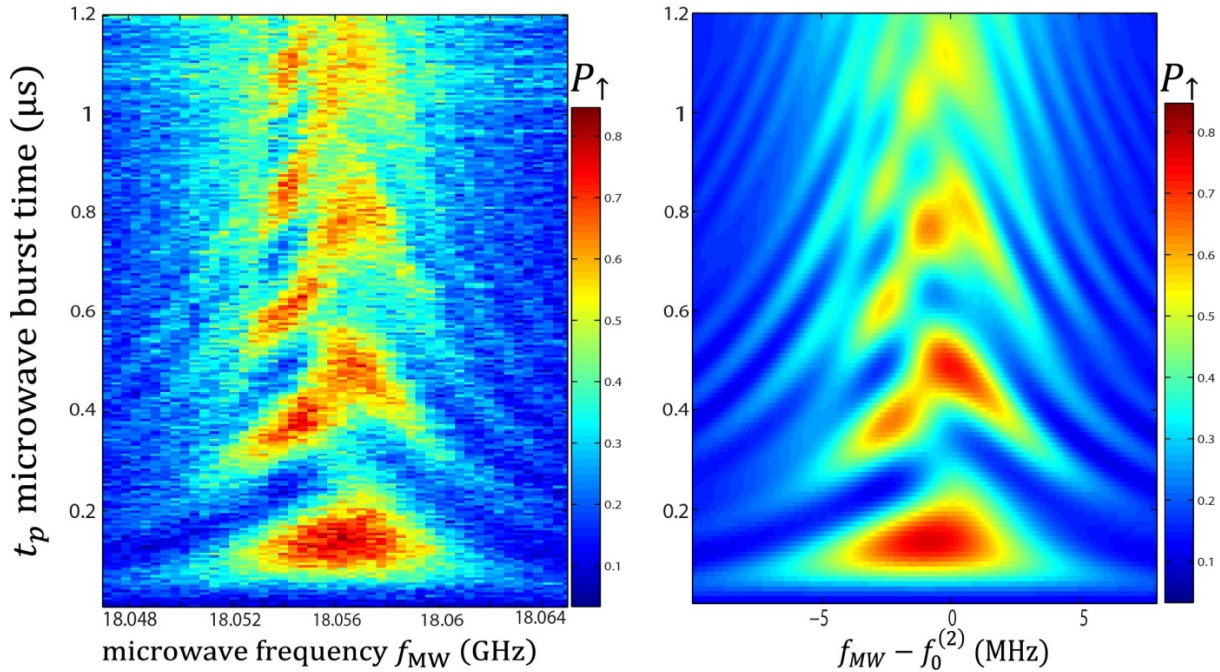

Fig. S4 Comparison of the data to a simulation for Rabi oscillations at $B_{ext} = 763.287$ mT. (**A**) The measured spin-up probability $P_\uparrow$ for a Rabi experiment. (**B**) The simulated spin-up probability $P_\uparrow$ using population fractions $\varepsilon^{(1)}:\varepsilon^{(2)} = 0.3:0.7$, Larmor linewidth $\sigma_{f_1} \sim 0.25$ MHz, readout fidelity parameters $\alpha = 6\%$, $\beta = 5\%$, $\gamma = 4\%$, and the two Rabi frequencies $f_1^{(2)} = 3.1$ MHz and $f_1^{(1)} = 4.1$ MHz.

## S5. Population and time dependence of the two resonances

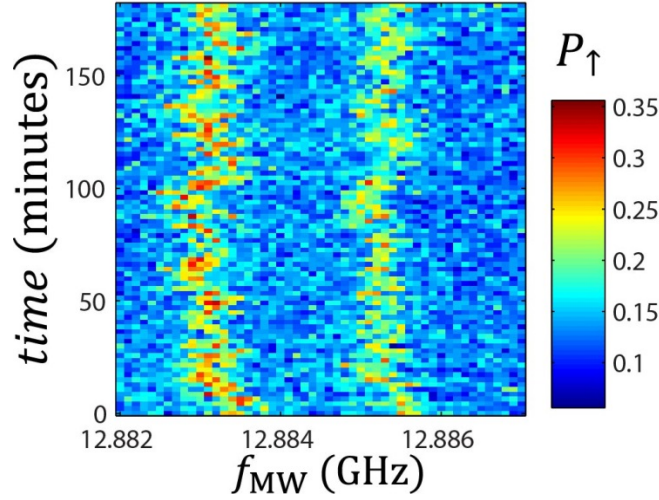

Fig. S5 Measured spin-up probability $P_\uparrow$ as a function of applied microwave drive frequency $f_{MW}$ and time (external field $B_{ext} = 560.783$ mT, power P = -33 dBm, microwave pulse duration $t_p = 700$ μs).

The raw data on which Fig. 2b is based are shown in Fig. S5. Each horizontal scan in the figure takes ~2 minutes (200 cycles, which takes 2s, per datapoint), and the scan is repeated 86 times. Fig. 2b shows the average of the 86 horizontal scans.

Fig. 2b shows the average of the 86 traces shown in Fig. S5. We see from Fig. S5 that the center of the resonance frequency $f_0$ fluctuates over time and the fluctuation behaviour is the same for the two resonances.

For the measurement of Fig. 2b, the applied microwave power is very low ($f_1 \ll \sigma_f$) and the burst time is very long ($t_p \gg T_2^*$).

Assuming that the ratio of the Rabi frequencies between two resonances is $\kappa$ ($f_1^{(1)} = \kappa\eta$, $f_1^{(2)} = \eta$) and we can neglect the unknown spread in $f_1$ for low microwave power, the ratio of the steady state spin flip probability at the two Larmor frequencies is

$$r(\eta) = \frac{P^{(1)}_{\uparrow\downarrow}\left(f_{MW}=f_0^{(1)}, t_p \gg T_2^*\right)}{P^{(2)}_{\uparrow\downarrow}\left(f_{MW}=f_0^{(2)}, t_p \gg T_2^*\right)} = \frac{\int df\, G(f) p_{\uparrow\downarrow}^{(1)}\left(f_1^{(1)}=\kappa\eta, f, f_{MW}=f_0^{(1)}, t_p \gg T_2^*\right)}{\int df\, G(f) p_{\uparrow\downarrow}^{(2)}\left(f_1^{(2)}=\eta, f, f_{MW}=f_0^{(2)}, t_p \gg T_2^*\right)},$$

where $p_{\uparrow\downarrow}$ and $G(f)$ are defined as in Eq. (S1) and Eq. (S2).

In the limit of low microwave power, the ratio $r(\eta)$ converges to:

$$r_0 = \lim_{\eta \to 0} r(\eta) = \kappa^2. \qquad \text{Eq.(S3)}$$

If we assume that the ratio of the Rabi frequencies between two resonances at low MW power is the same as the raio at high MW power determined in Section 8 of the supplementary material, then $\kappa = 1.53 \pm 0.19$.

The ratio of the measured peak amplitudes in Fig. 2b is 1.4±0.3 and it is the product of the ratio of the spin flip probabilities and the ratio of the populations:

$$r_0 \times \frac{\varepsilon^{(1)}}{\varepsilon^{(2)}} = 1.4 \pm 0.3. \qquad \text{Eq.(S4)}$$

From Eq. (S3) and Eq. (S4), we get

$$\frac{\varepsilon^{(1)}}{\varepsilon^{(2)}} = \frac{1.4 \pm 0.3}{\kappa^2} = \frac{1.4 \pm 0.3}{2.34 \pm 0.58} = 0.60 \pm 0.28.$$

From this relation, we get $\varepsilon^{(1)} : \varepsilon^{(2)} \sim 0.37 : 0.63$, consistent with the rough estimate of 0.3:0.7 based on the Rabi oscillations (see Section 4 of the supplementary information).

## S6. π pulse fidelity

The fidelity for flipping a spin using a π pulse is given in Eq. (S2). For the lower transition, using the values $f_1^{(1)} = 5$ MHz, $\sigma_f = 0.268$ MHz and $\sigma_{f_1} = 0.25$ MHz, we find

$$P^{(1)}_{\uparrow\downarrow}\left(f_{MW} = f_0^{(1)}, t_p = \frac{1}{2f_1^{(1)}} = 100\text{ns}\right) = \int df_1 \int df\, G^{(1)}(f) g^{(1)}(f_1) p^{(1)}_{\uparrow\downarrow}(f_1, f, f_{MW}, t_p)$$
$$= 0.99.$$

For the higher transition, using $f_1^{(2)} = 3.1$ MHz, $\sigma_f = 0.268$ MHz and $\sigma_{f_1} = 0.25$ MHz, we find

$$P^{(2)}_{\uparrow\downarrow}\left(f_{MW} = f_0^{(2)}, t_p = \frac{1}{2f_1^{(2)}} = 160\text{ns}\right) = \int df_1 \int df\, G^{(2)}(f) g^{(2)}(f_1) p^{(2)}_{\uparrow\downarrow}(f_1, f, f_{MW}, t_p)$$
$$= 0.97.$$

When we have a $\varepsilon^{(1)}: \varepsilon^{(2)} = 0.3:0.7$ contribution of the two resonances, $P_{\uparrow\downarrow}(= 0.3 \times P_{\uparrow\downarrow}^{(1)} + 0.7 \times P_{\uparrow\downarrow}^{(2)})$ reaches its maximum 0.79 when $P_{\uparrow\downarrow}^{(1)} = 0.53$ and $P_{\uparrow\downarrow}^{(2)} = 0.9$ at $t_p = 130ns$ and for $f_{MW} = f_0^{(2)}$.

# S7. Initialization fidelity and readout fidelity

For applications in quantum information processing it is important to know the read-out and initialization fidelities. These fidelities are characterized by three parameters, $\alpha$, $\beta$ and $\gamma$.
The parameter $\alpha$ corresponds to the probability that the sensing dot current exceeds the threshold even though the electron was actually spin-down, for instance due to thermally activated tunnelling or electrical noise.
The parameter $\beta$ corresponds to the probability that the sensing dot current does not cross the threshold even though the electron was actually spin-up at the end of the microwave burst time. The measurement time ($< 4$ ms) we used is much shorter than $T_1$ and so $\beta$ is not affected by $T_1$ decay (see supplementary section 3). It is limited by the bandwidth of the sensing dot current measurement (~20 kHz). $(1 - \beta)$ can be directly measured as the probability that the step of the electron jumping in during the initialization stage is missed using the same threshold value as is used for detection of the electron jumping out during the read-out stage. We find $\beta \sim 5\%$ (Fig. S6).

The parameter $\gamma$ corresponds to the probability that the electron is in spin-up instead of spin-down at the end of the initialization stage.
The measured spin-up probability $P_\uparrow$ can be written as follows using the parameters $\alpha$, $\beta$ and $\gamma$ and the probability for flipping the spin during manipulation, $P_{\uparrow\downarrow}$:

$$P_\uparrow = P_{\uparrow\downarrow}(1-\gamma)(1-\beta) + (1-P_{\uparrow\downarrow})\gamma(1-\beta) + (1-P_{\uparrow\downarrow})(1-\gamma)\alpha + P_{\uparrow\downarrow}\gamma\alpha \quad \text{Eq.(S5)}$$

When $P_{\uparrow\downarrow} = 0$ (i.e. the microwaves are applied far off-resonance or not at all), the measured spin-up probability can be expressed as follows:

$$P_\uparrow(P_{\uparrow\downarrow} = 0) = (1-\gamma)\alpha + \gamma(1-\beta) \quad \text{Eq.(S6)}$$
$$= \alpha + \gamma(1-(\alpha+\beta)) \quad \text{Eq.(S7)}$$

$P_\uparrow(P_{\uparrow\downarrow} = 0) \sim 10\%$ is measured (Fig. S6B. From this and Eq.(S7), we get an upper bound on $\alpha$.

$$P_\uparrow(P_{\uparrow\downarrow} = 0) > \alpha$$
$$10\% > \alpha \quad \text{Eq.(S8)}$$

As discussed above, $(1-\beta) \sim 95\%$ is measured. From this and Eq.(S7), we get an upper bound on $\gamma$.

$$P_\uparrow(P_{\uparrow\downarrow} = 0) > \gamma(1-\beta)$$
$$\frac{P_\uparrow(P_{\uparrow\downarrow} = 0)}{(1-\beta)} > \gamma$$
$$11\% > \gamma \quad \text{Eq.(S9)}$$

By looking at Fig. 3a, $P_\uparrow(f_{MW} = f_0^{(2)})$ reaches its maximum ~ 0.72 when $t_p \sim 130$ $ns$. Here, since $P_{\uparrow\downarrow}$ is expected to be large, the 2$^{nd}$, 3$^{rd}$ and 4$^{th}$ terms of Eq.(S5) are much smaller than the 1$^{st}$ term (each of them contains two factors much smaller than 1, whereas the 1$^{st}$ term contains no such small factors). So $P_\uparrow$ can be well approximated as follows.

$$P_\uparrow \sim P_{\uparrow\downarrow}(1-\gamma)(1-\beta) = 0.72$$
$$P_{\uparrow\downarrow}(1-\gamma) = \frac{0.72}{(1-\beta)} = 0.76. \qquad \text{Eq.(S10)}$$

Using the upper bound of $\gamma$ (Eq.(S9)), we can put bounds on $P_{\uparrow\downarrow}(f_{MW} = f''_0, t_p \sim 130ns)$:

$$0.76 < P_{\uparrow\downarrow}(f_{MW} = f''_0, t_p \sim 130ns) < 0.85. \qquad \text{Eq.(S11)}$$

Numerical simulation for $\sigma_{f_1} = 0.25\ MHz$ gives $P^{(2)}{}_{\uparrow\downarrow}(f_{MW} = f_0^{(2)}, t_p = 130ns) = 0.9$ and $P^{(1)}{}_{\uparrow\downarrow}(f_{MW} = f_0^{(2)}, t_p = 130ns) = 0.53$, where we note that the 130 ns burst time is longer respectively shorter than the burst time for a $\pi$ pulse for the lower and higher energy resonance. Then, using $\varepsilon^{(1)}:\varepsilon^{(2)} = 0.3: 0.7$, we obtain $P_{\uparrow\downarrow}(f_{MW} = f_0^{(2)}, t_p = 130ns) = 0.53 \times 0.3 + 0.9 \times 0.7 = 0.79$, which is consistent with Eq.(S11). Now, using $P_{\uparrow\downarrow}(f_{MW} = f_0^{(2)}, t_p = 130ns) = 0.79$ and Eq.(S11) we can estimate $\gamma = 4\%$.

Then, from Eq.(S6), we can also extract $\alpha$.

$$P_\uparrow(P_{\uparrow\downarrow} = 0) = (1-\gamma)\alpha + \gamma(1-\beta) = 10\%$$
$$0.96\alpha + 0.04 * 0.95 = 10\%$$
$$\alpha = 6\%.$$

We use $\alpha = 6\%$, $\beta = 5\%$, and $\gamma = 4\%$ in Eq.(S5) to compute the spin-up probability $P_\uparrow$ that can be expected in the measurement, which is shown in Fig. S3B, Fig. S4B and Fig. S7B.

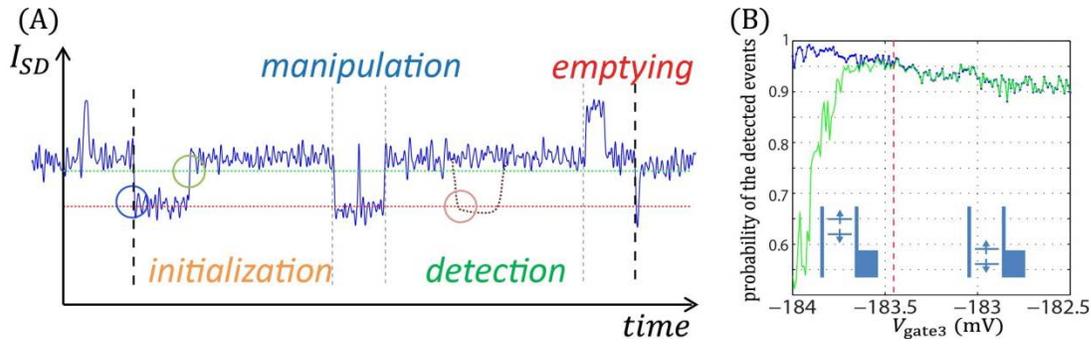

Fig. S6 Measurement of fidelity parameter $(1-\beta)$
(**A**) An example real time trace of the sensing dot current. The black dashed lines indicate the start and end of one cycle. When the recorded current dips below the threshold level indicated by the dotted red line during the detection stage, we conclude an electron tunnelled out from the dot to the reservoir. In this case, we infer the electron was spin up. When the signal remains above the threshold, we conclude the electron was spin down (the lowest energy spin state). (**B**) Blue trace: Measured probability that the sensing dot current passes below the threshold indicated by the red dotted line in panel (A) during the initialization stage, as a function of $V_{\text{gate3}}$ (averaged over 1000 cycles). Since the dot is always emptied during the previous stage, ideally we would always see the signal dip below the red threshold at the start of the initialization stage (blue circle in (A)). However, because of the finite bandwidth of the measurement, the dip will be missed if it is too fast. This occurs with the same probability as the probability for missing dips in the detection stage, and is thus a good measure of $(1-\beta)$. Green trace: Measured probability that the current subsequently passes the green threshold from below during the initialization stage, as a function of $V_{\text{gate3}}$ (green circle in (A)). When the green and blue traces coincide, the dot is filled during the initialization stage. When the dot level is high ($V_{\text{gate3}}$ is low, see also the schematic in the inset), the time it takes for an electron to tunnel in is long, and so $(1-\beta)$ is high, but the dot is not always filled (the green line is low here). As the dot level is lowered ($V_{\text{gate3}}$ is raised, see also inset), the tunnel rate increases, and the dot is always initialized, at the cost of a slightly lower value of $(1-\beta)$, due to the finite measurement bandwidth. The vertical red dashed line indicates the operating point used in the experiments.

## S8. Power dependence of the Rabi frequency

Fig. S6A shows the measured Rabi frequencies for the two resonance conditions, $f_1^{(1)}$ and $f_1^{(2)}$, as a function of the microwave amplitude emitted from the source. $f_1^{(1)}$ and $f_1^{(2)}$ are determined by the fast Fourier transform (FFT) of Rabi oscillations, as in Fig. 3b. The error bars arise from the finite number of points in the FFT. The linear fits shows that the ratio of the Rabi frequencies of two resonance transitions is $f_1^{(1)}/f_1^{(2)} = 1.53 \pm 0.19$.

Fig. S6B shows Rabi oscillations for a range of microwave amplitudes emitted from the source. The scattering and the low spin-up probability around microwave amplitude 500 mV~ 800mV may be due to a background charge switch that caused the dot to move away from the electrochemical potential alignment that is best for read-out. The measurement of Fig. S6B took 20 hours.

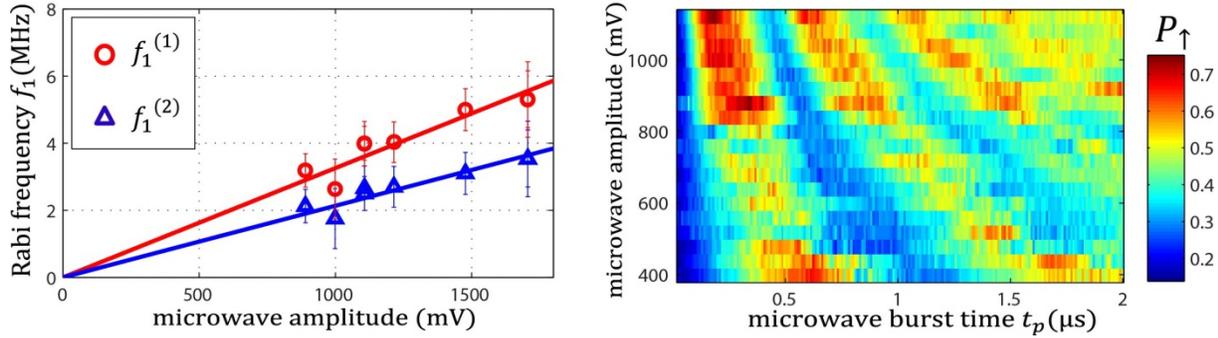

Fig. S7 Rabi oscillations versus microwave amplitude
(**A**) Rabi frequencies $f_1^{(1)}$ (red circles) and $f_1^{(2)}$ (blue circles) at $B_{ext} = 560.783$mT as a function of the microwave amplitude emitted from the source, as verified with a spectrum analyser. The solid lines are linear fits to the data. As expected, the Rabi frequency is linear in the driving amplitude. (**B**) The measured spin-up probability $P_\uparrow$ as a function of the microwave burst time $t_p$ and the microwave amplitude emitted from the source by applying microwave excitation at $f_{MW} = f_0^{(2)} = 12.885$ GHz. ($B_{ext} = 560.783$mT) We see the pattern expected for Rabi oscillations.

## S9. Measurements of Ramsey fringes

Here we give results of numerical simulations corresponding to the two-pulse Ramsey interference measurements of Fig. 3c. The overall procedure is analogous to that used for the simulations of the Rabi oscillations. Instead of a single microwave burst, we now have two bursts of duration $t_p = \frac{1}{4f_1^{(1)}}$ separated by a wait time $\tau$. The expression that the spin is flipped at the end of this sequence is as follows (same symbols as in the Rabi simulations, see Eq. (S1)[12]:

$$p_{\uparrow\downarrow}(f_1, f, f_{MW}, \tau) =$$
$$4\sin^2\theta \sin^2\left(\pi t_p \sqrt{(f_{MW}-f)^2 + f_1^2}\right)\left[\cos(\pi(f_{MW}-f)\tau)\cos\left(\pi t_p \sqrt{(f_{MW}-f)^2 + f_1^2}\right) - \cos\theta\sin(\pi(f_{MW}-f)\tau)\sin\left(\pi t_p \sqrt{(f_{MW}-f)^2 + f_1^2}\right)\right]^2$$
$$\text{with } \sin\theta = \frac{f_1}{\sqrt{(f_{MW}-f)^2 + f_1^2}}.$$

Here we can neglect the spread in $f_1$ because $t_p$ is short and its effect is small. Then the spin flip probability averaged over the Larmor frequency distribution is expressed as

$$P_{\uparrow\downarrow}(f_{MW}, t_p) = \int df \left(G^{(1)}(f)\varepsilon^{(1)}p^{(1)}{}_\uparrow + G^{(2)}(f)\varepsilon^{(2)}p^{(2)}{}_\uparrow\right). \qquad \text{Eq.(S12)}$$

Using Eq. (S10), $\alpha = 6\%$, $\beta = 5\%$, and $\gamma = 4\%$ in Eq. (S3), we compute the expected spin-up probability $P_\uparrow$ at the end of the Ramsey sequence, see Fig. S7(B). The corresponding data is shown in Fig. S7(A).

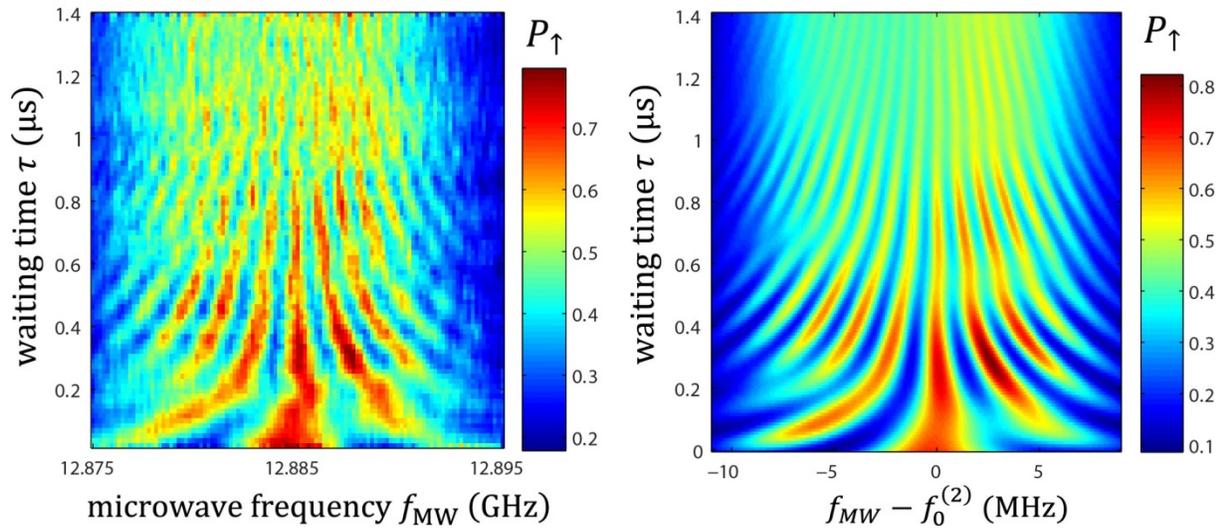

Fig. S8 Comparison of the data to a simulation for Ramsey fringes.
(**A**) The measured spin-up probability $P_\uparrow$ for a two-pulse Ramsey style experiment. The data are those shown in Fig. 3c but taking a moving average along $t_p$ over 5 points (79 ns). (**B**) The simulated spin-up probability $P_\uparrow$ using $\varepsilon^{(1)}:\varepsilon^{(2)} = 0.3:0.7$, $\sigma_f = 0.268$ MHz, $\alpha = 6\%$, $\beta = 5\%$, and $\gamma = 4\%$ as a function of $\tau$ and $f_{MW} - f_0^{(2)}$, also taking a moving average over 79 ns. There is good agreement between the data and simulation.

# S10. Difference in *g*-factors and Rabi frequencies between the two resonances

Here we discuss several possible explanations for the existence of two closely spaced electron spin resonance conditions, characterized by *g*-factors that differ by 0.015% and Rabi frequencies that differ by 50%.

As stated in the main text, we attribute the presence of two spin resonance signals to a partial occupation of the two lowest valley states. We can estimate the valley splitting $E_V$ from the 30/70 relative contributions of the two resonances (see main text), assuming it results from thermal equilibration between the two valley states. This gives $E_V \sim 0.85\, k_B T_e$, which for $T_e$ = 150 mK yields $E_V$ = 11 µeV. We note that the electron temperature may be somewhat larger since we apply microwave excitation to the sample, so the valley splitting may be larger as well. We have identified two mechanisms that can explain a 0.015% relative difference in the electron *g*-factors between the two valleys, defined as $2(g^{(2)}-g^{(1)})/(g^{(2)}+g^{(1)})$. The first is valley-dependent *g*-factor renormalization due to the transverse gradient magnetic field; the other is valley-dependent penetration of the electron wavefunction into the SiGe barrier region. We first discuss these two mechanisms. We then mention other potential mechanisms that cannot explain the observed *g*-factor shift.

(1) Tokura et al.,[43] find that the unperturbed Zeeman splitting $E_{0z}$ is renormalized in the presence of a magnetic field gradient as follows to a value $E_Z$ given by $E_z = E_{0z}\left[1 - \frac{0.5 M_{2,1}^2}{\Delta_{2,1}^2 - E_{0z}^2}\right]$. Here $M_{2,1} \sim g\mu_B (dB_\perp/dx) L$ is the perturbation matrix element between the ground orbital state with spin up and the first excited orbital state with spin down, and $L$ is the dot diameter. We see that $E_z$ depends on the orbital energy splitting (energy level spacing to the first excited state), $\Delta_{2,1}$. At lowest order in the valley-orbit coupling, $\Delta_{2,1}$ depends only on the orbital energy splitting, which can differ for the two valley states due to valley-orbit coupling [14]. Contributions to the renormalization of the g-factor from differences in the lateral positions of the different valley states can also occur, but are higher order in the valley-orbit coupling.
The difference in *g*-factors between the two valleys could then be explained if the two valley states exhibit sufficiently different orbital splittings $\Delta_{2,1}$. Valley-dependent orbital splitting arises from the valley-orbit interaction due to disorder at the interface, and can have important effects. In Ref.[15], it is estimated that the centers of the charge distributions of the two valley states can be separated by as much as the dot diameter, and differences in orbital splitting between the two valleys can be 20% or more. Taking $dB_\perp/dx = 1 mT/nm$, $E_{0z} \sim$ 60 µeV, $\Delta_{2,1}^{(1)} \sim$ 400 µeV and $\Delta_{2,1}^{(2)} \sim$ 320 µeV (where the superscripts refer to the two resonances as in the main text), we obtain corrections to $E_{0z}$ of 0.013% and 0.010%. The difference between the two corresponds to a difference in g-factors of 0.003%, within a factor of 5 of the observed value.
A valley-dependent orbital splitting can also account for the observed difference in Rabi frequencies for the two resonances. From Ref. 43,46, neglecting the contribution from spin-orbit interaction as it is small in Si/SiGe, we roughly have that $\omega_{Rabi} = \frac{g\mu_B}{2h} eE_{a.c.}\left|\frac{dB_\perp}{dx}\right|\frac{L^2}{\Delta}$, where $E_{ac}$ is the a.c. electric field generated by the nearby gate. Given that $L \propto 1/\sqrt{\Delta}$, it follows that $\omega_{Rabi} \propto 1/\Delta^2$. Then, we have that $\omega_{Rabi}^{(1)}/\omega_{Rabi}^{(2)} = (\Delta^{(2)}/\Delta^{(1)})^2$. Assuming that $E_{ac}$ is equal for the two valley states, the factor 1.5 between the Rabi frequencies of the two resonances can be explained by a ~20% difference in orbital level spacing, $2(\Delta^{(1)} - \Delta^{(2)})/(\Delta^{(1)} + \Delta^{(2)})$. This is consistent with the difference in orbital splitting needed to explain the *g*-factor shifts.

(2) A second explanation for the *g*-factor shifts could be that the two valley states penetrate differently into the SiGe barrier. This effect also gives rise to valley splitting. For *g*-factors, the state with the largest probability in the barrier should have the *g*-factor closest to SiGe. It is difficult to estimate the resulting *g*-factor shift because the *g*-factors in SiGe alloys are not well known. Our rough estimate yields a *g*-factor shift of 0.0025%, which is 6 times smaller than the experiment, but is still comparable. We view this mechanism as less likely than mechanism (1) above because observing the difference in Rabi frequencies would require that the different valley states have significantly different direction of wavefunction motion. In principle, further experiments have the potential to distinguish these two mechanisms for *g*-factor shifts. Valley-dependent penetration should be similar in similar devices, and its dependence on extrinsic parameters (e.g., accumulation gate voltages) should be systematic. On the other hand, valley-orbit renormalization should vary significantly from device to device.

We now briefly consider explanations for the g-factor shifts that yield less successful agreement with experiment.

(3) In principle, the combination of valley-orbit coupling and spin-orbit coupling could give rise to valley-dependent *g*-factor shifts. The renormalization in the *g*-factor from this mechanism is proportional to the inverse square of the spin-orbit length[17]. According to Ref.[18], the spin-orbit coupling strength in quantum well structures is three orders of magnitude smaller in Si than in III-V semiconductors. Since such *g*-factor renormalization effects are small already in GaAs, we can conclude that the change in g-factor mediated by this mechanism in Si will be much smaller than the 0.015% that is observed experimentally.

(4) As mentioned above, valley-orbit coupling may cause a lateral separation of the centers of the charge distributions for the two valley states[15]. When this effect is combined with local fluctuations of the Ge concentration in the SiGe alloy, it yields slightly different *g*-factor shifts for the two states. In general, the *g*-factor shift described in (2) (above) would be expected to dominate over such a disorder effect. However, because valley-orbit coupling depends on the interference between valley states[14], destructive interference could suppress the dominant *g*-factor shift in (2). Our simulations (not reported here) indicate that it is possible for the disorder-induced effect to dominate, though still smaller than the estimate given in (2) above.

(5) Finally, we consider explanations for the two closely spaced spin resonance conditions that do not invoke valley physics. A natural thought is that we may be driving spin transitions in a two- or three-electron manifold, either in a single dot or in a double dot. Under appropriate conditions, this could give rise to closely spaced spin resonance frequencies with *g*-factors around 2. However, in this scenario, whenever microwave excitation is applied at either one of the two resonance frequencies, spin transitions would be induced 100% of the time. In the experiment, in contrast, when applying microwave excitations resonant with the lower (upper) resonance frequency, there is a contribution to the signal only ~30% (70%) of the time. If the dot location jumped between two positions, for instance due to a background charge that is hopping back and forth, a 30/70 occupation would be possible. Due to the magnetic field gradient, we can also expect different spin splittings for different dot locations. However, the difference in spin splittings would be a fixed value set by the stray field from the micromagnet (as soon as it is fully polarized). In contrast, in the measurements, the difference between the resonance frequencies varies linearly with magnetic field (Fig. 2a). We have not been able to come up with other plausible explanations except those related to valley physics presented above.